\documentclass[12pt]{article}
\usepackage{epsfig}
\usepackage{url}
\makeatletter
\def\alt{\mathrel{\mathpalette\gl@align<}}
\def\agt{\mathrel{\mathpalette\gl@align>}}
\def\gl@align#1#2{\lower.6ex\vbox{\baselineskip\z@skip\lineskip\z@
\ialign{$\m@th#1\hfil##\hfil$\crcr#2\crcr\sim\crcr}}}
\makeatother

\begin{document}
\begin{flushright}
MI-TH-1626\\
August, 2016
\end{flushright}
\vspace*{1.0cm}
\begin{center}
\baselineskip 20pt
{\Large\bf
Is Electroweak Symmetry Breaking  Still Natural \\ in the MSSM?
} \vspace{1cm}

{\large
Bhaskar Dutta$^1$ and Yukihiro Mimura$^2$}
\vspace{.5cm}

$^1${\it
Department of Physics, Texas A\&M University,
College Station, TX 77843-4242, USA
}\\
%\vspace{.5cm}
$^2${\it
Department of Physics, National Taiwan University, Taipei 10617, Taiwan
}
\\

\vspace{.5cm}
%$^\dagger${\it
%}\\

\end{center}

\begin{center}{\bf Abstract}\end{center}

The absence of any signal of supersymmetry (SUSY) at the LHC has raised  the SUSY particle mass
scale compared to $Z$ boson mass $M_Z$. %Is the electroweak symmetry breaking vacuum still natural in SUSY? 
We investigate the naturalness of the electroweak symmetry breaking after considering radiative symmetry breaking along with 125 GeV Higgs mass. We find that the important quantity to measure the naturalness of the hierarchy between the SUSY scale and $M_Z$
is the separation between the radiative symmetry breaking scale, i.e., where $m_{H_u}^2+\mu^2$ turns negative for large $\tan\beta$ case ($\mu$ is the Higgsino mass and $m_{H_u}$ is the SUSY breaking up-type Higgs boson mass)
and the average stop mass.
Using this measure, one can show that the  electroweak symmetry breaking can be natural even if  
$\mu$ is large contrary to the prevailing claim that  $\mu$ is needed to be small to maintain the naturalness.
\thispagestyle{empty}

\bigskip
\newpage

\addtocounter{page}{-1}

\section{Introduction}
\baselineskip 18pt

The minimal supersymmetric standard model (MSSM)
has been a great candidate of the models beyond the standard model,
which can solve the hierarchy problem between the weak scale 
and the GUT/Planck scale.
The standard lore is that SUSY is confronted with a fine-tuning problem 
due to the experimental results at the Large Hadron Collider (LHC).
In fact, the electroweak symmetry breaking condition implies
that the $Z$ boson mass and the SUSY breaking masses
are roughly of the same order unless there is a cancellation.
Therefore,  the SUSY breaking particles
are considered to be around several hundred GeV.
Using this optimistic expectation,
people have  discussed the predictions for rare decay observables,
such as $B_s \to \mu^+\mu^-$,
which can be enhanced to be much larger than the standard model (SM).
However, the LHC results so far have shown no deviation from the SM prediction.
The direct searches also have raised the bounds of the gluino mass to be more than 1.9 TeV 
at ATLAS \cite{ATLAS:2016kts} and 1.6 TeV at CMS \cite{CMS:2016mwj}.
Surely, all these different experimental results are consistent with each other.
On the other hand, in order to realize the observed Higgs boson mass to be 125 GeV in the MSSM,
a heavy gluino is more preferable implicitly.
However, the problem of having   heavy colored SUSY particles is the fine-tuning 
in the electroweak symmetry breaking condition.
In the scenario of the minimal number of the SUSY breaking parameters,
we need 0.1\% level cancelation ($O(1000)-O(1000)\sim 1$) in the tree-level relation.
This is often called as a little hierarchy problem,
since the cancellation
is much better compared to the SM which has quadratic divergence
in the Higgs boson mass, $O(M_P^2)-O(M_P^2) \sim O(M_W^2)$.
The cancellation is needed between the SUSY breaking Higgs mass
and a supersymmetric mass (Higgsino mass $\mu$) in the superpotential.
Such a model which gives a smaller size of $\mu$ is often described
as {\it natural} SUSY \cite{Giusti:1998gz,Kitano:2006gv,Papucci:2011wy,Baer:2012uy,Kim:2016rsd},
and the parameter space which gives a large size $\mu$ is {\it unnatural}.

The Higgs mechanism in the SM has an open question:
Why is the squared Higgs mass negative and how is the electroweak symmetry broken?
MSSM can answer the question since the squared Higgs mass becomes negative radiatively \cite{Inoue:1982pi}.
%The idea of the radiative symmetry breaking can also explain the large hierarchy 
%between the weak scale and the Planck scale.
%
Actually, if the Higgsino and SUSY breaking masses are of the same order by a mechanism \cite{Giudice:1988yz},
the electroweak symmetry breaking condition can be naturally satisfied
at a scale much lower than the cutoff scale. 
The remaining question is why the SUSY breaking mass scale is close to the scale
where the symmetry breaking condition becomes satisfied radiatively.
About a decade ago, the authors discussed 
a measure of the sensitivity quantity in the {\it radiative} symmetry breaking,
and the probability distribution of the little hierarchy \cite{Dutta:2007az}.
The little hierarchy is found to be probable among the electroweak symmetry vacua.
At that time, however, we did not take the Higgs boson mass to be 125 GeV (which may need a large trilinear stop coupling) into account and  the current LHC bounds on the SUSY particles makes the little hierarchy to be less probable as long as gaugino mass unification is assumed.

In this paper, we first  discuss a sensitivity quantity which is suitable for the
radiative electroweak symmetry breaking. We then show that,
if the gaugino unification is not assumed, the little hierarchy indicated by the LHC 
is still probable by using the sensitivity quantity.
We claim that
the important quantity to measure the naturalness of the little hierarchy 
is the separation between the radiative symmetry breaking scale
and the average stop mass.
Using the measure, one can say that the radiative breaking can be natural even if the 
Higgsino mass is large.
In order to highlight the importance of this type of parameter space, 
we call it as {\it naturally unnatural} electroweak symmetry breaking.
The naturalness arguments should depend on how to measure the sensitivity,
and we should understand the little hierarchy from different points of view.
We  question the belief that the naturalness in the electroweak symmetry breaking 
is measured
only by the tree-level relation without taking the running of the mass parameters into account.

This paper is organized as follows: in section 2, 
we discuss a measure for naturalness of the electroweak vacuum, in section 3, 
we discuss the naturalness in the context of $m_h$=125 GeV, in section 4,
 we discuss the phenomenological consequences of a natural electroweak vacuum and we conclude in section 5.

\section{Developing a Measure for  Naturalness of the electroweak vacuum}

Minimizing the tree-level Higgs potential 
\begin{equation}
V = m_1^2 v_d^2 + m_2^2 v_u^2 - (m_3^2 v_d v_u + c.c)
+ \frac{g_2^2+ g^{\prime 2}}{8} (v_d^2 - v_u^2)^2,
\end{equation}
by the Higgs vacuum expectation values (VEVs) ($v_d = \langle H_d^0 \rangle$
and $v_u = \langle H_u^0 \rangle$),
we obtain
\begin{equation}
\frac{M_Z^2}{2} = \frac{m_1^2 - m_2^2 \tan^2\beta}{\tan^2\beta -1},
\qquad
\sin2\beta = \frac{2m_3^2}{m_1^2 + m_2^2},
\end{equation}
where $\tan\beta = v_u/v_d$.
The quadratic mass terms are given by SUSY breaking Higgs masses, $m^2_{H_d}$ and $m^2_{H_d}$,
Higgsino mass $\mu$, and SUSY breaking bilinear Higgs mass $B\mu$:
\begin{equation}
m_1^2 = m_{H_d}^2 + \mu^2, \quad
m_2^2 = m_{H_u}^2 + \mu^2, \quad
m_3^2 = B\mu.
\end{equation}
Without any loss of generality, we can make $m_3^2$  to be real.
The minimization condition can provide the relation between the $Z$ boson mass
and the Higgs mass parameters.
The relation can be rewritten as
\begin{equation}
M_Z^2 = \frac{4((m_3^2)^2 - m_1^2 m_2^2)}{(m_2^2-m_1^2)\cos2\beta + (m_1^2+m_2^2) \cos^22\beta}.
\end{equation}
If $m_1^2 > m_2^2$, one obtains $\tan\beta >1$, and thus, $(m_2^2 - m_1^2) \cos2\beta >0$.
The condition of the electroweak symmetry breaking at the tree-level is
\begin{equation}
(m_3^2)^2 > m_1^2 m_2^2.
\end{equation}
One also needs a stability condition 
\begin{equation}
m_1^2 + m_2^2 > 2 |m_3^2|.
\end{equation}

%In SUSY limit, the symmetry breaking condition is not satisfied.
Even if the condition is not satisfied (namely, $m_1^2 m_2^2 - (m_3^2)^2  > 0$) at a ultraviolet scale,
the symmetry breaking can happen via the renormalization group equation (RGE) evolutions of SUSY breaking mass parameters
due to a large top quark Yukawa coupling.
Because the mass parameters run by RGEs, 
the tree-level relation depends on the scale $Q$
\begin{equation}
M_Z^2 (Q) \equiv 2\frac{m_1^2(Q) - m_2^2 (Q) \tan^2\beta(Q)}{\tan^2\beta(Q) - 1}.
\end{equation}
It is important to know the scale where the tree-level relation is evaluated.
Indeed, the improved potential contains a loop correction, and the 1-loop correction can be written as \cite{Coleman:1973jx}
\begin{equation}
\Delta V = \frac{1}{64\pi^2} \sum_J (-1)^{2J} (2J+1) m_J^4 \left( \ln \frac{m_J^2}{Q^2} - \frac32 \right),
\end{equation}
where $J$ stands for a spin of the fields.
Since the mass $m_J$ depends on the Higgs VEVs, the minimization condition of the total potential 
has to include the derivatives of the loop correction $\partial \Delta V/\partial v_{u,d}$.
In principle, the total potential does not depends on the scale $Q$ (up to wave function renormalization),
and thus, one can use a scheme that the 
scale $Q$ is chosen to make the two derivatives to be small \cite{Casas:1998cf}.
Naming the scale when the two derivatives are small as $Q_S$,
\begin{equation}
\frac{\partial \Delta V}{\partial v_u} (Q=Q_S) \approx
\frac{\partial \Delta V}{\partial v_d} (Q=Q_S)\approx
0,
\end{equation}
we find that the tree-level relation is approximately correct for $Q=Q_S$:
\begin{equation}
M_Z^2 \simeq M_Z^2 (Q_S).
\end{equation}
%
% both two derivatives by $v_u$ and $v_d$ are not necessarily small simultaneously 
One can find that the scale $Q_S$ is roughly the geometrical average of the stop masses,
$\sqrt{m_{\tilde t_1} m_{\tilde t_2}}$.
Surely, the symmetry breaking condition has to be satisfied
at $Q=Q_S$.
As described, the electroweak symmetry breaking happens radiatively,
which means that there is a scale that the condition $(m_3^2)^2 - m_1^2 m_2^2 > 0$ becomes to be satisfied.
We call the scale $Q_0$ as a symmetry breaking scale (namely, $M_Z^2(Q_0)=0$):
\begin{equation}
(m_3^2(Q_0))^2 - m_1^2 (Q_0)m_2^2(Q_0) = 0.
\end{equation}
The stability condition should be also satisfied and we call the scale $Q_{st}$
as 
\begin{equation}
m_1^2(Q_{st}) + m_2^2(Q_{st}) - 2|m_3^2(Q_{st})| = 0.
\end{equation}
The radiative electroweak symmetry breaking
condition can be expressed as
\begin{equation}
Q_{st} < Q_S < Q_0.
\end{equation}

The tree-level relation tells us that the $Z$ boson mass is directly related to the
quadratic mass parameters in the Higgs potential, and thus, it is naively of the same order as the 
SUSY breaking mass scale.
If the $Z$ boson mass scale is much smaller than the typical SUSY breaking masses,
an adjustment of the parameters is needed,
and the evaluation scale $Q_S$ can be taken to be close to the symmetry breaking scale $Q_0$.
The statement can be interpreted as follows:
By expanding the scale-dependent tree-level function $M_Z^2(Q)$ around the scale $Q_0$, we obtain
\begin{equation}
M_Z^2 \simeq M_Z^2 (Q_S) = \left. \frac{d M_Z^2}{d\ln Q}\right|_{Q=Q_S} \ln \frac{Q_S}{Q_0} + \cdots,
\end{equation}
where $M_Z^2 (Q_0) = 0$ by definition, 
and therefore, we obtain
\begin{equation}
M_Z^2 \cos^22\beta \simeq 
\left(
\frac{dm_2^2}{d\ln Q} \sin^2 \beta + 
\frac{dm_1^2}{d\ln Q} \cos^2 \beta -
\frac{dm_3^2}{d\ln Q }\sin2\beta
\right) \ln \left(\frac{Q_0}{Q_S}\right)^2.
\end{equation}
The RGEs of $m_i^2$ relates the quantities $\frac{dm_i^2}{d\ln Q}$
to the SUSY breaking mass parameters.
Indeed, the radiative electroweak symmetry breaking is triggered by
the quantity $\frac{dm_2^2}{d\ln Q}$,
and if $\frac{dm_2^2}{d\ln Q}$ is larger, $\ln Q_0/Q_S$ has to be smaller.

For a large $\tan\beta$ ($\tan\beta \agt 5$), one can consider the following approximate relations:
\begin{eqnarray}
M_Z^2 &\simeq& - 2 m_2^2 (Q_S), \\
M_Z^2 &\simeq& \frac{dm_2^2}{d\ln Q} \ln \left(\frac{Q_0}{Q_S}\right)^2.
\label{MZ}
\end{eqnarray}
The 1-loop RGE of $m_2^2 = m_{H_u}^2 + \mu^2$ is given as
\begin{eqnarray}
\frac{dm_{H_u}^2}{d\ln Q} &=& \frac{1}{8\pi^2} \left(3 (y_t^2 (m_{\tilde t_L}^2 + m_{\tilde t_R}^2 + m_{H_u}^2) + A_t^2) - (g^{\prime2} M_1^2 + 3 g^2 M_2^2) + \frac12 g^{\prime2} S \right), \\
\frac{d\mu^2}{d\ln Q} &=& \frac{1}{8\pi^2}(3 y_t^2 + 3 y_b^2 + y_\tau^2 - g^{\prime2} - 3 g_2^2) \mu^2,
\end{eqnarray}
where $S$ is a trace of all the scalar masses with hypercharge weight.
We obtain 
\begin{equation}
M_Z^2 \simeq
\frac{1}{8\pi^2} \left(
3 y_t^2 (m_{\tilde t_L}^2 + m_{\tilde t_R}^2) + 3A_t^2
 - g^{\prime2} (M_1^2 + \mu^2) -3 g_2^2 (M_2^2 + \mu^2)  + \frac12 g^{\prime2} S
\right) \ln \left(\frac{Q_0}{Q_S}\right)^2.
\end{equation}
where $y_t^2 (m^2_{H_u} + \mu^2) = - y_t^2 M_Z^2/2$ and $y_{b,\tau}^2 \mu^2$ terms are neglected.
If one of the stop masses and/or $A_t$ are more than about 2 TeV
(which is implied by %by the direct search at the LHC and 
the Higgs mass to be $125$ GeV \cite{Ellis:1990nz,Carena:2011aa}), 
one requires a small $\ln Q_0/Q_S$ naively.
Namely, the evaluation scale $Q_S$ ($\simeq \sqrt{ m_{\tilde t_1} m_{\tilde t_2} }$) is very close to the symmetry breaking scale $Q_0$.

We remark that the properties of the scales $Q_0$ and $Q_S$,
which are important to describe the little hierarchy between the SUSY breaking masses 
and $Z$ boson mass.
The symmetry breaking scale $Q_0$ is determined
if the boundary conditions of RGEs (for SUSY breaking mass parameters
and Higgsino mass $\mu$) are fixed at the GUT/Planck scale.
The RGEs are homogeneous equations, and thus, the
scale $Q_0$ does not depend on the overall rescaling of the dimensionful parameters.
Therefore, if 
all the SUSY breaking parameters and the Higgsino mass are parameterized by a scale parameter
$M_S$ as
\begin{equation}
m_{\tilde t_{L,R}}^2 =  \hat m_{\tilde t_{L,R}}^2 M_S^2,\
A_t = \hat A_t M_S,\
M_{1,2,3} = \hat M_{1,2,3} M_S,\
B = \hat B M_S,\
\mu = \hat \mu M_S,
\label{hatted}
\end{equation}
and so on,
the scale $Q_0$ does not depend on $M_S$,
but depends on the dimensionless hatted parameters.
On the other hand,
$Q_S$ surely depends on $M_S$ directly.
Therefore, it is useful to describe the relation using the parameters $M_S$ and $Q_0$.
We obtain
\begin{equation}
M_Z^2 \approx \alpha_1 M_S^2 \ln \frac{Q_0}{\alpha_2 M_S},
\end{equation}
where $\alpha_1$ and $\alpha_2$ are dimensionless coefficients for the hatted parameters.

The sensitivity quantity of a function $f(x)$ is defined as \cite{Ellis:1986yg}
\begin{equation}
\Delta[f(x)] \equiv \left|\frac{d \ln f}{d \ln x}\right|^{-1}.
\end{equation}
The sensitivity for the $M_S$ and $Q_0$ can be calculated as
\begin{equation}
\Delta [M_Z(M_S)] \approx \frac{\ln \left(\frac{Q_0}{Q_S}\right)^2}{\left|1-\ln \left(\frac{Q_0}{Q_S}\right)^2\right|},
\qquad
\Delta [M_Z(Q_0)] \approx \ln \left(\frac{Q_0}{Q_S}\right)^2.
\end{equation}
Therefore, as expected,
the sensitivity can be specified by $\ln{Q_0}/{Q_S}$.
Precisely speaking, the coefficients $\alpha_1$ and $\alpha_2$ depends on $M_S$ by RGEs, 
but we neglect the higher order dependence.
We note that the $Z$ boson mass is insensitive to $M_S$,
if $\ln Q_0/Q_S \simeq 1/2$.
This is nothing but the solution of no-scale supergravity model,
in which the scale $M_S$ is determined by the minimization of the 
electroweak Higgs potential \cite{Ellis:1983sf,Barbieri:2000kj,Dutta:2007xr}.

We insist that the size of $\ln Q_0/Q_S$ 
is a good quantity to specify the fine-tuning 
of the {\it radiative} electroweak symmetry breaking.
Since we fix the $Z$ boson mass,
the size of $\ln Q_0/Q_S$ directly relates to the size of $dm_2^2/d\ln Q$.
Several comments are needed for this demand:%insistence:
\begin{enumerate}

\item
The fine-tuning quantity is usually characterized by
\begin{equation}
\Delta [M_Z(\mu)] = \frac{M_Z^2}{2\mu^2},
\end{equation}
from the tree-level relation (fixing $M_S$):
\begin{equation}
\frac{M_Z^2}{2} = \frac{m_1^2 - m_2^2 \tan^2\beta}{\tan^2\beta -1}=
-\mu^2 + \frac{m_{H_d}^2 - m_{H_u}^2 \tan^2\beta}{\tan^2\beta -1}.
\end{equation}
This selection of the quantity is valid because the cancellation between $\mu^2$ and
$(m_{H_d}^2 - m_{H_u}^2 \tan^2\beta)/(\tan^2\beta -1)$ is needed
(if $|m_{H_u}^2|$ is large).
As explained,
\begin{equation}
\mu^2=
\frac{m_{H_d}^2 - m_{H_u}^2 \tan^2\beta}{\tan^2\beta -1}
\end{equation}
satisfies
at the scale $Q_0$  by definition.
Therefore, the cancellation between them is equivalent to the closeness of $Q_0$ and $Q_S$.

\item
In the radiative breaking scenario in MSSM, 
$\mu^2$ and $(m_{H_d}^2 - m_{H_u}^2 \tan^2\beta)/(\tan^2\beta -1) \simeq -m_{H_u}^2$
intersect at a lower scale $Q_0$.
The scenario nicely explains the hierarchy between the GUT/Planck scale
and the electroweak symmetry breaking scale.
As explained, the scale is determined by a set of 
hatted parameters in Eq.(\ref{hatted}) in the boundary conditions.
For a realistic purpose, the scale $Q_0$ has to be chosen to be at the TeV scale.
The scale $Q_0$ is sensitive to the boundary conditions.
For example,
the sensitivity quantity can be obtained as
\begin{equation}
\Delta [Q_0(\mu)] \approx \frac{1}{2\mu^2} \frac{dm_2^2}{d\ln Q}.
\end{equation}
Surely, this is sensitive to $\mu$ for a large value of $\mu$.
By definition,
we have 
$\Delta[M_Z(\mu)] = \Delta[M_Z(Q_0)] \Delta[Q_0(\mu)]$.
We should address this sensitivity quantity $\Delta[Q_0(\mu)]$ %distinctly 
separately to discuss the fine-tuning
in the radiative electroweak symmetry breaking.

\item
The issue of the little hierarchy is
the hierarchy between a typical SUSY breaking mass (stop masses)
and $Z$ boson mass.
A fine-tuning sensitivity in our vacuum
among the electroweak symmetry vacua in MSSM should be addressed.
In the landscape picture of the electroweak symmetry breaking,
the symmetry breaking scale is not necessarily the TeV scale.
The  sensitivity quantity $\Delta[M_Z(\mu)]$, which is typically used, describes the
tuning to adjust the scale $Q_0$ at a TeV scale.
Indeed, though the radiative breaking scenario nicely explains the
hierarchy between the scale $Q_0$ and the GUT/Planck scale,
the scale $Q_0$ is surely sensitive to the boundary condition,
and fixing $Q_0$ at a scale contains a fine-tuning. %
We, therefore, suggest to use the
fine-tuning index $\Delta[M_Z(Q_0)] = \Delta[M_Z(\mu)]/\Delta[Q_0(\mu)]$
by removing 
the sensitiveness of $Q_0$
in order to discuss the fine-tuning of the electroweak symmetry breaking.
%($\Delta[Q_0(\mu)]$).
%

If the typical SUSY breaking scale $Q_S$ is fixed in the landscape of 
radiative electroweak symmetry breaking vacua,
it is valid to discuss the sensitivity using the quantity by $\Delta[M_Z(\mu)]$ or $\Delta[Q_0(\mu)]$.
It is probable that there are various values of $Q_S$ in the landscape,
and thus we should employ the quotient $\Delta[M_Z(\mu)]/\Delta[Q_0(\mu)]$ to discuss the naturalness
of our living vacuum in the landscape.

The aim of this paper is to discuss the electroweak symmetry breaking vacua where $\Delta[M_Z(Q_0)]$ is large (i.e., the vacua are natural),
but $\Delta[M_Z(\mu)]$ and $\Delta[Q_0(\mu)]$ are small (i.e. the vacua look unnatural).
We call them {\it naturally unnatural electroweak symmetry breaking vacua}.

\item
We do not know the probability distribution function of the overall scale $M_S$ in the landscape.
Therefore, it is not quite valid to discuss only the sensitivity quantity.
Suppose that the overall scale is equally probable (i.e. distribution function $D[M_S] = $ const),
and then, the distribution function of $q=\ln Q_0/M_S$ can be obtained as\footnote{
We remark that the scale $Q_0$ is sensitive to the hatted parameter (up to overall scale)
but it does not depend on the overall scale.
}
\begin{equation}
D[q] \equiv D[M_S] \frac{dM_S}{dq} \propto e^{-q}.
\end{equation}
This means that the smaller value of $\ln Q_0/M_S$ is more probable.
Namely, although the radiative electroweak symmetry breaking does not occur in the most of the vacua,
a little hierarchy is probable in the landscape of electroweak symmetry breaking \cite{Dutta:2007az,Giudice:2006sn}.

\item
As explained,
in the MSSM, the symmetry breaking condition is satisfied by RGE evolution,
and the cancellation between $\mu^2$ and $(m_{H_u}^2 - m_{H_u}^2 \tan^2\beta)/(\tan^2\beta-1) \simeq -m_{H_u}^2$
happens at a scale $Q_0$.
Therefore, we stress that the tuning condition for the little hierarchy can be
interpreted as the closeness of $Q_0$ and the evaluation scale $Q_S$ where
the tree-level minimization relation is approximately correct.
In such a point of view, 
the size of the Higgsino mass parameter is not necessarily specify the
fine-tuning in the landscape of electroweak symmetry breaking
since the intersection of $\mu^2$ and $-m_{H_u}^2$ is irrespective of the size of $\mu$.
Rather the size of $dm_2^2/d\ln Q$ is more important
due to Eq.(\ref{MZ}).
Indeed, 
if $dm_2^2/d\ln Q$ is smaller,
the $Z$ boson mass $M_Z$ is less sensitive to $Q_0$ and $Q_S$,
but $Q_0$ is more sensitive to $\mu$.
In the Hyperbolic/Focus point solution \cite{Chan:1997bi,Feng:1999mn} (if $Q_S$ is less than the focus point), 
the Higgsino mass $\mu$ is smaller and ``tree-level natural" symmetry breaking vacua is obtained.
In the solution, however, $dm_2^2/d\ln Q$ is large and a fine-tuning is needed to bring $Q_0$ and 
$Q_S$ to be very close.
One can intuitively understand that
 the size of $|m_{H_u}^2|$ easily becomes larger if $dm^2_2/d\ln Q$ is large,
 and one needs an adjustment to make $|m_{H_u}^2|$  small at the evaluation scale.
We need to be aware of the fact  that the sensitivity can depend on how the parameter space is sliced
to discuss the fine-tuning of the electroweak symmetry breaking.

\end{enumerate}

\section{Naturalness of electroweak vacuum in MSSM with 125 GeV Higgs mass}

As explained in the previous section, in the radiative electroweak symmetry breaking scenario, 
the symmetry breaking condition, $m_1^2 m_2^2 < (m_3^2)^2$, is satisfied by RGE evolution,
and the
$Z$ boson mass can be written approximately as 
\begin{equation}
M_Z^2 \simeq \frac{dm_2^2}{d\ln Q} \ln \left(\frac{Q_0}{Q_S}\right)^2,
\end{equation}
for a large $\tan\beta$.
The RGE for the up-type Higgs squared mass $m_2^2$ can be written  
\begin{equation}
\frac{dm_2^2}{d\ln Q} \simeq \frac{1}{8\pi^2} \left(
3 y_t^2 (m_{\tilde t_L}^2 + m_{\tilde t_R}^2) + 3A_t^2
 - g^{\prime2} (M_1^2 + \mu^2) -3 g_2^2 (M_2^2 + \mu^2) 
\right),
\label{RGE}
\end{equation}
where we omit the trace of scalar masses with the hypercharge weight factors  for simplicity to show.
The RGE  depends on the SUSY breaking mass parameters,
and thus, the $Z$ boson mass is naively the scale of the SUSY breaking masses 
(which are typically stop masses)
multiplied by a loop factor.
Therefore, in the radiative breaking scenario, the naive SUSY breaking scale should lie around a TeV scale.
If SUSY breaking masses are larger, one needs a fine-tuning between the
symmetry breaking scale $Q_0$ and the evaluation scale $Q_S$.
%
%In fact, the symmetry breaking scale $Q_0$ does not depend on the overall scale $M_S$
%because the RGEs are homogenous equations, while $Q_S$ directly depends on $M_S$.
%Therefore, the closeness of those two scales are mystery in the electroweak symmetry breaking
%in MSSM if the SUSY particles are heavy (as implied by the LHC experiments).

Let us assume  that the gaugino masses are unified at the GUT scale.
The RGE solution of the gaugino masses at 1-loop is
\begin{equation}
\frac{M_3}{\alpha_3} = \frac{M_2}{\alpha_2} = \frac{M_1}{\alpha_1} = {\rm const}.,
\end{equation}
and the gluino mass $M_3$ is heavier than the others at the low energy scale.
The gluino masses pushes up the SUSY breaking masses of colored particles.
Therefore, 
the contributions from $M_1$ and $M_2$ in Eq.(\ref{RGE}) become negligible.
In that case, neglecting the negative contribution in the equation,
we obtain that the $Z$ boson mass is given approximately as
\begin{equation}
M_Z^2 \simeq \frac{3}{8\pi^2} \left(y_t^2 (m_{\tilde t_L}^2+ m_{\tilde t_R}^2) + A_t^2\right) 
\ln \left(\frac{Q_0}{Q_S}\right)^2.
\end{equation}
The sensitivity quantity $\Delta[M_Z(Q_0)]$ is determined by the stop mass parameters:
\begin{equation}
\Delta[M_Z(Q_0)] \simeq \ln\left(\frac{Q_0}{Q_S}\right)^2 \simeq \frac{8\pi^2 M_Z^2}{3(y_t^2(m_{\tilde t_L}^2+ m_{\tilde t_R}^2) + A_t^2)}.
\end{equation}
As we have already mentioned that
in order to obtain 125 GeV Higgs mass, this quantity has to be small 
($\alt0.01$)\footnote{Compared to the quantity $\Delta[M_Z(\mu)]$, the sensitivity quantity is larger due to the loop factor.}
and the two scales $Q_0$ and $Q_S$ have to be very close.

Can we relax the sensitivity in the MSSM? The answer is yes.
In order to do that, the gaugino unification condition should be broken and the $M_1^2$ and $M_2^2$ terms 
are made to be comparable to the positive terms in Eq.(\ref{RGE}).
In fact, if the positive terms $3(y_t^2 (m_{\tilde t_L}^2+ m_{\tilde t_R}^2) + A_t^2 )$
and the negative terms $- g^{\prime2} (M_1^2 + \mu^2) -3 g_2^2 (M_2^2 + \mu^2) $
are canceled at the weak scale, the sensitivity quantity $\Delta[M_Z(Q_0)]$ can become large.
One may think that the cancellation calls for another fine-tuning.
However, the stop mass parameters, in principle, increase at lower energy by RGE because they have color,
while $M_1$, $M_2$ decrease because $SU(2)_L\times U(1)_Y$ is not  asymptotically free.
As a consequence, the cancellation happens naturally somewhere at low energy
if $dm_2^2/d\ln Q$ is negative at high scale as a boundary condition.
Surely, one needs some adjustments to make 
the scale where $m_2^2$ becomes stationary
lie nearly at the weak scale.
%and if one only calculate the sensitivity quantity $\Delta[Q_0(\mu)]$ or $\Delta[M_Z(\mu)]$, it may look unnatural.
%
It is intuitively obvious that $Q_0$ is sensitive to $\mu$ if $-m_{H_u}^2$ and $\mu^2$ intersect
at the scale where $m_{H_u}^2$ becomes stationary. 
Nevertheless, the quantity $\Delta[M_Z(Q_0)]\simeq \ln(Q_0/Q_S)^2$ can be large enough
to say that it is the natural radiative electroweak symmetry breaking, namely,
the two scales $Q_0$ and $Q_S$ are not very close.

\begin{figure}[t]
\begin{center}
\includegraphics[width=0.48\textwidth]{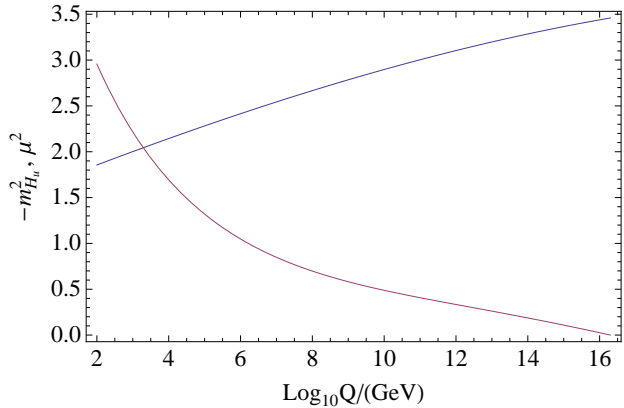}
\caption{
The RGE running of $-m_{H_u}^2$ (red) and $\mu^2$ (blue) in the case of $M_1=M_2=M_3$ at GUT scale.
As explained in the text, the shape of the running dose not depend on the overall scale factor, 
and we choose %a unit as
 $M_3=1$ as a unit  at the  GUT scale. A value of $\mu/M_3$ is chosen to make
the two lines to cross at a few TeV.
}
\end{center}
\label{run1}
\end{figure}

\begin{figure}[t]
\begin{center}
\includegraphics[width=0.48\textwidth]{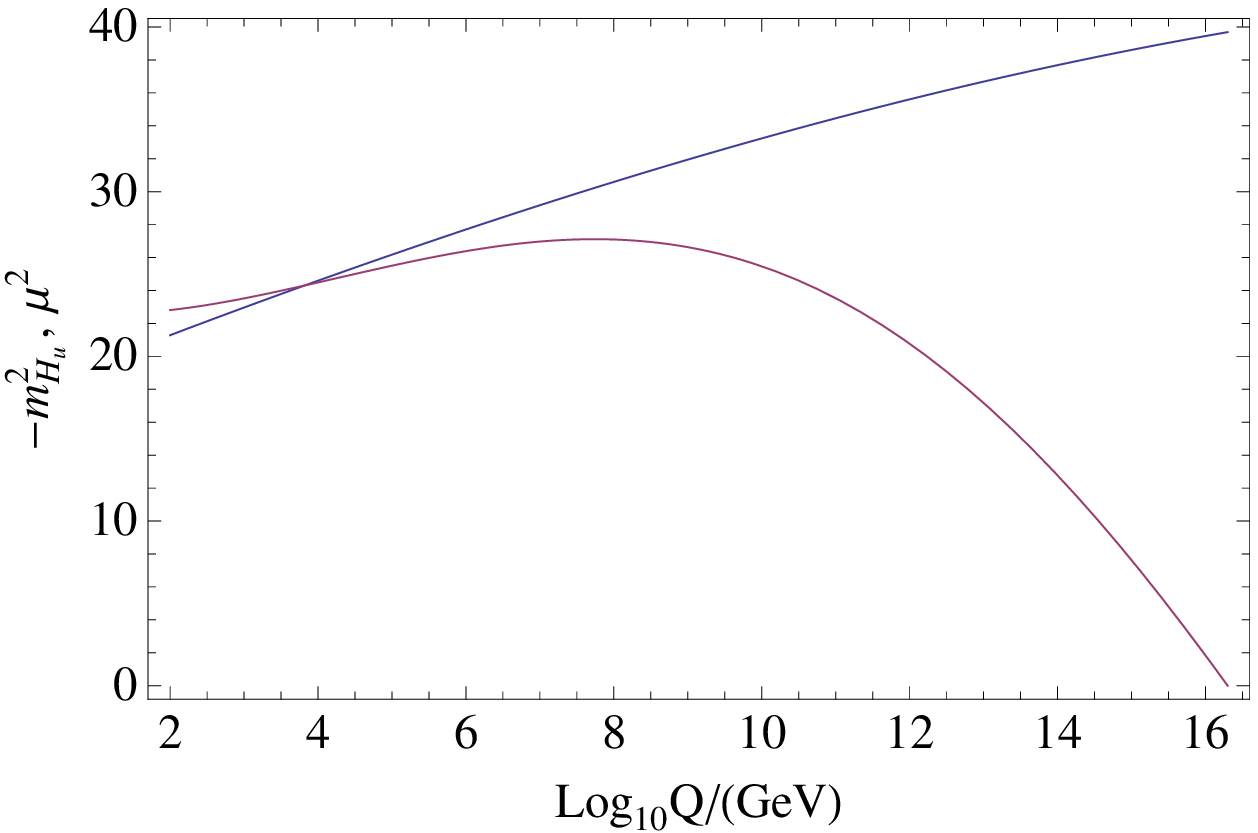}
\includegraphics[width=0.48\textwidth]{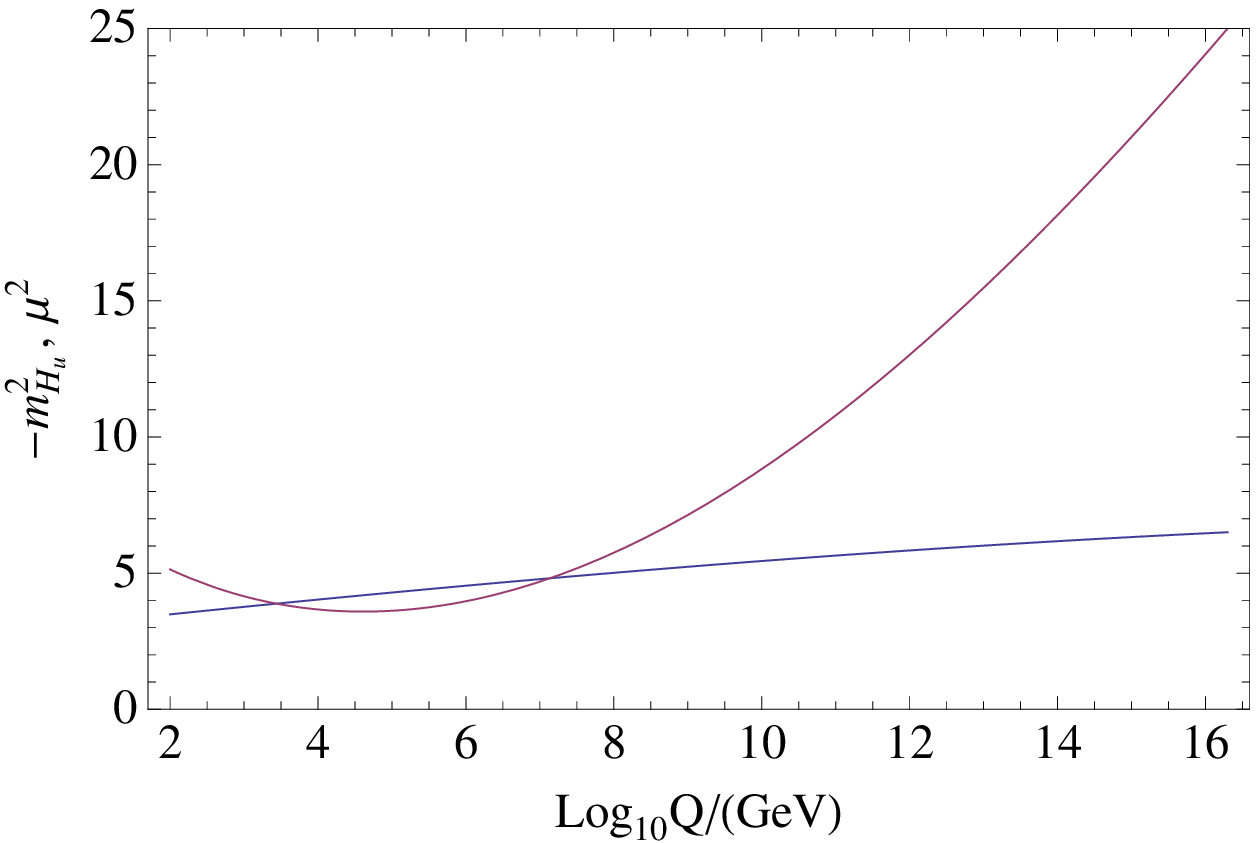}
\caption{
The RGE running of $-m_{H_u}^2$ and $\mu^2$ without assuming the gaugino mass unification.
We choose the parameters to make the running of $m_{H_u}^2$ stationary near the weak scale.
}
\end{center}
\label{run2}
\end{figure}

Let us study the RGE running of $-m_{H_u}^2$ and $\mu^2$ in concrete examples.
As we noted, the RGEs are homogeneous differential equations, and the running of the squared masses
can be rescaled without changing the shape.
We take $M_3 = 1$ (at the unification scale) as a normalization.
In Fig.1, the running of $-m_{H_u}^2$ and $\mu^2$ is shown under the gaugino unification
condition $M_1=M_2=M_3$.
The SUSY breaking masses of squarks and sleptons, $m_{H_u}^2$ and $m_{H_d}^2$ are chosen to be
zero at the unification scale in this example.
Then, $-m_{H_u}^2$ runs in the conventional way and the behavior of the running at the TeV scale
is governed by the stop mass parameters as described above.
In Fig.2, the examples are shown where  $\frac{dm_{H_u}^2}{d\ln Q}$ becomes zero below (left) and above (right)
the weak scale by choosing the SUSY breaking parameters.
As explained, it can naturally happen by breaking the gaugino mass unification condition.
In these examples, we choose $M_2/10 \sim M_3 = M_1$.
As one can see, $\frac{dm_2^2}{d\ln Q}$ is small even though the size of $\mu^2$ is larger than the one in
Fig.1,
and thus the sensitivity quantity $\Delta[M_Z(Q_0)] \simeq \ln(Q_0/Q_S)^2$ is larger.
We note that the case where the stationary point of $-m_{H_u}^2$ is above the weak scale,
one needs to choose $m_{H_u}^2$ to be negative at the high scale as a boundary condition.
Surely, it can be done within the parameter space of MSSM, or one can construct a model to create such a condition.

The RGE evolutions are fully described by dimensionless parameters up to an overall scale parameter $M_S$.
In order to adjust the $Z$ boson mass to be 91 GeV, one has to choose $M_S$.
As explained, the sensitivity quantity $\Delta[M_Z(M_S)]$ is larger for smaller $\frac{dm_2^2}{d\ln Q}$
even if the Higgsino mass $\mu$ is large.
One can learn  the argument of this parameter sensitivity of the adjustment in a practical way which is as follows.
In Figs.1 and 2, the $\mu$ parameter (more precisely $\mu/M_3$ in the examples)
is chosen to make $Q_0$ (the scale where $-m_{H_u}^2$ and $\mu^2$ intersects) 
to be about 2-3 TeV.
If one wants to adjust the $Z$ boson mass with fixed SUSY particle mass (gluino, squarks, etc.),
one also needs to tune $Q_0$.
While the scale $Q_0$ does not depend on the overall scale $M_S$,
the tuning of $Q_0$ is sensitive to $\mu/M_3$ especially if $dm_{H_u}^2/d\ln Q$ is small as one can see from the figures.
After all, one can argue that the $Z$ boson mass is sensitive to $\mu$ parameter.
This argument can be expressed by an identity:
\begin{equation}
\Delta[M_Z(\mu)] = \Delta[M_Z(Q_0)] \Delta[Q_0(\mu)].
\end{equation}
It is true that the $Z$ boson mass is sensitive to the $\mu$ parameter
if $\mu$ is large.
However, to argue the naturalness of the radiative electroweak symmetry breaking,
one should see how the SUSY breaking Higgs mass parameter runs.
In the radiative breaking scenario, the scale $Q_0$ is surely sensitive to $\mu/M_3$ and other dimensionless parameters.
We suggest that the quotient $\Delta[M_Z(Q_0)] = \Delta[M_Z(\mu)]/\Delta[Q_0(\mu)]$ is
suitable to characterize the naturalness in the little hierarchy,
instead of just referring the tree-level relation.

\section{Phenomenological implication}

Although the particle spectrum depends on the detail of the boundary conditions,
the low energy spectrum is constrained
because of the large sensitivity quantity $\Delta[M_Z(Q_0)]$.
In this section, we study the low energy spectrum from the constraint
in the naturally unnatural electroweak symmetry breaking.

As described in the previous section,
we expect that the wino and/or bino are heavy 
to make $\frac{d m_2^2}{d\ln Q}$ small so that the RGE evolution is stationary near TeV scale. 
However, since  one of the wino and bino can be light,
 the lightest of them  can be the dark matter candidate.
Because the gaugino mass unification is not supposed, 
the lightest SUSY particle mass is free to choose theoretically even if the gluino mass is fixed.

The size of the gluino mass is important for the RGE running of the colored SUSY particle masses.
The RGE of $m_{H_u}^2$ does not depend on the gluino mass,
but it depends on the stop masses and $\frac{d m_2^2}{d\ln Q}$ cannot be kept to be small
if gluino is heavy.
In other words, the curvature of the RGE running, which is neglected in the previous sections,
becomes important for a heavier gluino mass.
For a heavy gluino, thus, the scale where $\frac{d m_2^2}{d\ln Q}$ becomes zero and the scale $Q_0$ needs to be close which requires another tuning.
As a consequence, we expect that the gluino mass is less than several TeV. 
%even in the scenario.

Let us estimate the gluino mass from the stationary condition of $m_2^2$ near TeV scale.
Suppose that the wino mass balances the positive contributions from the stop masses.
Then, we obtain
\begin{equation}
g_2^2 M_2^2 \sim y_t^2 (m_{\tilde t_L}^2 + m_{\tilde t_R}^2) + A_t^2.
\end{equation}
If the arithmetic mean of the stop squared masses is $(2\ {\rm TeV})^2$ and the trilinear coupling is chosen to obtain 125 GeV Higgs boson mass,
the wino mass is roughly 10 TeV.
Defining the ratio $M_2/M_3= k$ at GUT scale,
we obtain the gluino and wino mass ratio (at TeV scale) as
$M_{\rm wino}/M_{\rm gluino} \simeq k/3$.
Therefore, 
the gluino mass can be roughly estimated
as
\begin{equation}
M_{\rm gluino} \sim \frac{30}{k} \ {\rm TeV}.
\end{equation}
According to our numerical study,
the SUSY breaking Higgs mass can be stationary near TeV scale when we choose $k \sim 5-O(10)$
(depending to the other SUSY breaking parameters).
As a consequence, 
in this naive estimate,
we expect that the gluino mass lies at several TeV.
We note that even if the mean stop mass is larger,
the estimated gluino mass is not proportionally larger
because the trilinear coupling can become smaller.
Even if the stop mass is 10 TeV, we estimate that the gluino mass is less than 10 TeV.

\section{Conclusions and Discussions}

%Human mind 
We sometimes make a wrong estimate
to measure the rarity of the coincidence of two parameters.
For a simple example, 
one can raise the rarity of rolling two dices and obtaining the same number.
It should be distinguished from the two dices being the same particular number,
and the rarity should be divided by the possible number of the dice.
Surely, it is an obvious example, but people often misestimate the rarity 
of the similar types of coincidence. %  daily life.
In a different kind of analogy of  fine-tuning in nature, 
the closeness of the apparent sizes of the sun and the moon is often discussed.
In order to think about the fine-tuning,
we should recall the fact that 
the moon is getting farther away from the earth at about 4 cm per year.
It is said that the tidal force due to the moon may be important to create life,
and the apparent size of the moon was bigger than the sun as a initial condition.
The rate of  increasing distance between the earth and the moon and the lifetime of the civilized intellectual creatures
should be taken into account when the coincidence in the apparent sizes is discussed. 
These analogies of course do not match with the naturalness in the electroweak
symmetry breaking completely, but they may be instructive.

The run-I and II data at the LHC so far have pushed up the bounds of gluino and/or squark masses.
Although it is still possible that one of the scalar top quarks is light and the trilinear scalar coupling
is large to obtain the 125 GeV Higgs mass in the MSSM and
there is a tuning in the (running) stop mass parameters with a heavy gluino mass
or in the tree-level electroweak symmetry breaking condition.
Before concluding that the tuning is unnatural and the low energy SUSY may be fictional, 
we should reconsider what is natural in the radiative electroweak symmetry breaking in the MSSM
since the fine-tuning in the SUSY models are still more natural compared to the cancellation required in the SM to solve the problem of the quadratic divergence of the Higgs mass.
In fact, the idea of the radiative electroweak symmetry breaking in the MSSM is quite natural since
the squared Higgs mass becomes negative radiatively due to the O(1) top Yukawa coupling
at a scale which is much lower than the cutoff scale.
This occurs due to  the fact that the SUSY breaking Higgs squared mass $m_{H_u}^2$ is driven  negative due to 
the loop diagram in which scalar top quarks propagate.
The SUSY contribution to the Higgs mass comes from so-called the $\mu$ parameter,
which does not run very rapidly.
Therefore, $-m_{H_u}^2$ and $\mu^2$ intersect at a lower energy scale (we call the scale as $Q_0$) 
and the total (SM-like) Higgs mass $m_{H_u}^2+\mu^2$ (for a large $\tan\beta$)
becomes negative radiatively.
The $Z$ boson mass is obtained $M_Z^2 = -2(m_{H_u}^2+\mu^2)$ at the tree-level,
which provides a good approximation if it is evaluated
at the scale $Q_S$ (which is roughly equal to a geometrical average of scalar top masses).
Usually, the cancellation between $-m_{H_u}^2$ and $\mu^2$ at $Q_S$ 
is used to discuss  naturalness of the little hierarchy.
The two mass parameters have to be very close if $|m_{H_u}^2|$ is large.

In this paper, we questioned the fine-tuning arguments which involve only $\mu^2$ 
and $-m_{H_u}^2$ without considering the running of them for the radiative electroweak symmetry breaking.
It is true that a small $\mu$ parameter provides a natural realization of the symmetry braking,
but we should not throw out another type of natural realization of the radiative breaking from the criteria of naturalness.
As explained in the text, the scale $Q_0$ does not change 
even if the overall mass scale parameter given in Eq.(\ref{hatted}) is turned to be a larger value.
The scale where the cancellation between $\mu^2$ and $-m_{H_u}^2$ happens can occur at $Q_S = Q_0$.
The cancellation between $\mu^2$ and $-m_{H_u}^2$ %at a particular scale 
contains the sensitivity of the scale
$Q_0$ which depends on $\mu/M_3$ and other dimensionless parameters.
We therefore suggest to divide the sensitivity quantity $\Delta[M_Z(\mu)]$ by 
the sensitivity of $Q_0$ ($\Delta[Q_0(\mu)]$) in order to obtain the sensitivity quantity 
of the radiative breaking.
The important quantity for describing  naturalness  can be interpreted as the separation of $Q_0$ and $Q_S$
(namely, $\ln Q_0/Q_S$).
The separation of  two scales depends on the running of the Higgs masses around the symmetry breaking scale $Q_0$.
The radiative electroweak symmetry breaking is natural when the scalar top quarks average scale is lighter than $Q_0$ and the running of the SUSY breaking Higgs mass becomes stationary near the TeV scale.
%
%The LHC bounds of the colored sparticle masses imply
%that the two scales $Q_0$ and $Q_S$ have to be close.
%
%We nay need another type of tuning between the symmetry breaking scale $Q_0$ and
%the scale where the running of the Higgs masses is stationary,
%and thus 
We found  that the SUSY particles should  be around less than several TeV.
We should therefore wait for more run-II data at the LHC without a prejudice from the naturalness argument  based on the cancellation between $\mu^2$
and $-m_{H_u}^2$.

\section*{Acknowledgments}

%\noindent

The work of  B. D.
is supported in part by the DOE grant DOE Grant DE-FG02-13ER42020. 
The work of Y.M. is supported by the Excellent Research Projects of
 National Taiwan University under grant number NTU-ERP-105R8915.

\end{document}